\documentstyle[aps,prb]{revtex}
\begin{document}

\title{Superconducting states in ferromagnetic metals} 
\author{V.P.Mineev}
\address{Commissariat a l'Energie Atomique, DSM, Departement de
Recherche Fondamentale sur la Matiere Condensee, SPSMS, 38054
Grenoble, France } 
\date{ Submitted 16 May 2002}
\maketitle
\begin{abstract}
The symmetry of the superconducting states arising directly from the
ferromagnetic states in crystals with cubic and orthorombic
symmetries is described.  The symmetry nodes in the quasiparticle
spectra of such states are pointed out if they exist.

The superconducting phase transition in the ferromagnet is accompanied
by the formation of superconducting domain structure consisting of complex
conjugate states imposed on the ferromagnet domain structure with the
opposite direction of the magnetization in the adjacent domains.  

The interplay between stimulation of a nonunitary
superconducting state by the ferromagnetic moment and supression of
superconductivity by the diamagnetic orbital currents is
established.

\end{abstract}
\bigskip

\section{Introduction}

A new class of superconducting materials has been revealed very
recently where the superconducting state appears from another ordered
state of the material - namely ferromagnetic state.  There are now
several metallic compounds demonstrating the coexistance of
superconductivity and itinerant ferromagnetism.  These are $UGe_{2}$
\cite {1,2}, $ZrZn_{2}$ \cite{3}, $URhGe$ \cite{4}.  The
superconducting states in these materials have to be preferably spin
triplet to avoid the large depairing influence of the exchange field.  Moreover
it seems reasonable \cite{5} that these are the states where only
electrons with the spin down direction of the spins are paired, as is
the case in the $A_{1}$-phase of superfluid He-3 \cite {6}.  Then the
interaction between ferromagnetic and Cooper pair magnetic moments will
stimulate the superconducting state.  The explanation of the phase
diagram of $ZrZn_{2}$ based on this idea has been proposed \cite {7}. 
At first sight it seems plausible becouse $ZrZn_{2}$ has a cubic
cristalline structure allowing multicomponent unconventional superconducting
states with spontaneous magnetization.  On the contrary the first 
discovered ferromagnet-superconductor $UGe_{2}$ has an orthorombic
structure.  The orthorombic point group obeys only one-dimensional
representations that prevents the formation of a superconducting state with
spontaneous magnetization in the crystals with strong spin-orbital
coupling as a result of a spontaneous phase transition from the normal
state \cite {8}.  However I.Fomin has recently \cite{9} shown that the
magnetic superconducting phases may arise from the normal ferromagnet
state even in the orthorombic crystal with strong spin-orbital
coupling .  It means that in this case the stimulation of the
superconductivity by the ferromagnetism also takes place.

The goal of this article is to present the detailed analysis of the problem of
interaction of triplet pairing superconductivity  with magnetization
in the ferromagnetic metals.  To investigate this problem one must first
have the symmetry classification scheme for the superconducting states
arising from the ferromagnetic normal state.  The point is that the
classification of unconventional superconducting states arising from
a nonmagnetic normal state, that has been established in the papers
\cite{10,11,12}, does not include the new ferromagnet - superconducting
states arising from the normal state with broken time reversal
symmetry.  So, the discussion of interplay of the stimulation of 
superconductivity by the ferromagnetism of itinerant electrons and the
supression of superconductivity by diamagnetic currents is forestalled
by the symmetry classification of possible triplet superconducting
states arising directly from a normal ferromagnet state in crystals
with an inversion centre.  All superconducting magnetic classes in the
crystals with orthorombic (Section 2) and cubic symmetry (Section 3)
are described and the corresponding superconducting order parameters
are presented.  The existing symmetry nodes in the spectra of the
elementary exitations are pointed out.  

It is shown that in the superconducting state the ferromagnetic domain structure
with opposite direction of magnetization in the adjacent domains
causes the appearence of the superconducting domain structure with the
complex conjugate order parameters and the opposite directions of the
Cooper pair magnetic moments.

\section{Superconductivity in the ferromagnet orthorombic metals}

\subsection{Superconducting states}

Let us consider first a ferromagnetic orthorombic crystal with
spontaneous magnetization along one of the symmetry axis of the second
order chosen as the $z$-direction.  The symmetry group 
\begin{equation}
    G=M \times U(1)
\label{1} 
\end{equation}
consists of the so called magnetic class\cite{13} $M$ and the group of
the gauge transformations $U(1)$.  Any magnetic superconducting state
arising directly from this normal state corresponds to the one of the
subgroups of the group $G$ characterized by broken gauge symmetry.  In
the given case $M$ is equal to $D_{2}(C_{2}^{z})=(E, C_{2}^{z},
RC_{2}^{x}, RC_{2}^{y})$, where $R$ is the time reversal operation. 
Let us look first on the subgroups of $G$ being isomorphic to the
initial magnetic group $D_2(C_2^{z}¥)$ and constructed by means of
combining  its elements with phase factor $e^{i\pi}¥$ being an
element of the group of the gauge transformations $U(1)$.  The
explicit form of these classes are
\begin{equation}
D_{2}(C_{2}^{z})=
(E, C_{2}^{z}, RC_{2}^{x}, RC_{2}^{y}),
\label{e2}
\end{equation}

\begin{equation}
\tilde D_{2}(C_{2}^{z})= (E, C_{2}^{z},
RC_{2}^{x}e^{i\pi},RC_{2}^{y}e^{i\pi}),
\label{e3}
\end{equation}

\begin{equation}
D_{2}(E)= (E, C_{2}^{z}e^{i\pi}, RC_{2}^{x}e^{i\pi},
RC_{2}^{y}),
\label{e4}
\end{equation}
\begin{equation}
\tilde D_{2}(E)= (E, C_{2}^{z}e^{i\pi}, RC_{2}^{x}, RC_{2}^{y}e^{i\pi}).
\label{e5}
\end{equation}

    The superconducting states are characterized by broken gauge symmetry. 
At the same time the phase transition from the normal paramagnetic
state with symmetry $D_{2}\times U(1)$ to the normal ferromagnetic
state with symmetry $D_{2}(C_{2}^{z})\times U(1)$ obeys this symmetry. 
That is why the phase transition from a normal paramagnetic state to a
normal ferromagnetic state and from a normal ferromagnetic state to a
superconducting ferromagnetic state can not have the same origine
contrary to the statement in the paper \cite{9}.  As a result the
corresponding phase transition lines may intersect each other only
accidentally in isolated points in the  $(P,T)$ plane.  In particular there
is no reason for the coincidence of these lines exactly in the quantum
critical point at T=0.  In the existing orthorombic compound $UGe_{2}$
the ferromagnetism and superconductivity at $T=0$ disappear
simultaneously above the critical pressure about 17.6 kbar, the low
temperature part of para-ferro phase transition near this pressure is
however of the first order \cite {2}.

 To each of the superconducting magnetic classes  corresponds an order
 parameter.  All these vector (triplet) order parameters in the crystal
 with inversion center and strong spin-orbital coupling have the form
\begin{equation}
{\bf d}({\bf R},{\bf k})=\eta({\bf R}){\bf \Psi}({\bf k}),
\label{e6}
\end{equation}
\begin{equation}
{\bf \Psi}({\bf k})=\hat {\bf x}f_{x}({\bf k})+ \hat {\bf 
y}f_{y}({\bf
k}) + \hat {\bf z}f_{z}({\bf k}) ,
\label{e7}
\end{equation}
where $\hat x, \hat y, \hat z$ are the unit vectors of the spin 
coordinate
system pinned to the crystal axes and $f_{x}({\bf k}), \ldots$ are the
odd functions of momentum directions of pairing particles on the 
Fermi 
surface.  Functions ${\bf \Psi}({\bf k})$ for each superconducting
state obey a normalization condition
\begin{equation}
\langle {\bf \Psi}^{*}({\bf k}){\bf \Psi}({\bf k})\rangle=1 ,
\label{e8}
\end{equation}
where the angular brackets denote the averaging over
${\bf k}$ directions .

The general form of the order parameters
for the states (\ref{e2})-(\ref{e5}) have been pointed out in the paper
\cite{9}.  We write them here in somethat different form:
\begin{equation}
\Psi^{A_{1}¥}({\bf k})= \hat {\bf 
x}(k_{x}u_{1}^{A_{1}¥}+ik_{y}u_{2}^{A_{1}¥})+ \hat
{\bf y}(k_{y}u_{3}^{A_{1}¥}+ik_{x}u_{4}^{A_{1}¥})+
\hat {\bf z}(k_{z}u_{5}^{A_{1}¥}+ik_{x}k_{y}k_{z}u_{6}^{A_{1}¥}),
\label{e9}
\end{equation}
\begin{equation}
\Psi^{A_{2}}({\bf k})= \hat {\bf
x}(ik_{x}u_{1}^{A_{2}}+k_{y}u_{2}^{A_{2}})+ \hat {\bf
y}(ik_{y}u_{3}^{A_{2}}+k_{x}u_{4}^{A_{2}})+ \hat {\bf
z}(ik_{z}u_{5}^{A_{2}}+k_{x}k_{y}k_{z}u_{6}^{A_{2}}),
\label{e10}
\end{equation}
\begin{equation}
\Psi^{B_{1}}({\bf k})= \hat {\bf x}(k_{z}u_{1}^{B_{1}}+i
k_{x}k_{y}k_{z}u_{2}^{B_{1}})+ \hat {\bf 
y}(ik_{z}u_{3}^{B_{1}}+
k_{x}k_{y}k_{z}u_{4}^{B_{1}})+ \hat {\bf 
z}(k_{x}u_{5}^{B_{1}}
+ik_{y}u_{6}^{B_{1}}),
\label{e11}
\end{equation}
\begin{equation}
\Psi^{B_{2}}({\bf k})= \hat {\bf x}(ik_{z}u_{1}^{B_{2}}+
k_{x}k_{y}k_{z}u_{2}^{B_{2}})+ \hat {\bf
y}(k_{z}u_{3}^{B_{2}}+i k_{x}k_{y}k_{z}u_{4}^{B_{2}})+ \hat
{\bf z}(ik_{x}u_{5}^{B_{2}} +k_{y}u_{6}^{B_{2}}),
\label{e12}
\end{equation}
where $u_{1}^{A}, \ldots$ are real functions of $k_{x}^{2}, 
k_{y}^{2},
k_{z}^{2}$.  It is worth noting that the state $\Psi^{A_{2}} $ transforms
as $i{\Psi^{A_{1}¥*}}$ and the state $\Psi^{B_{2}}$ transforms as
$i{\Psi^{B_{1}*}}$.

From the expressions for the order parameters (\ref{e9})-(\ref{e12}) one
can conclude that the states $A$ and $B$ have in general no symmetry
nodes in the quasiparticle spectrum.  Only occasional nodes appear for a
particular form of the functions $u_{1}^{A}, \ldots$.

  The classification of the states in quantum mechanics
corresponds to the general statement by E.Wigner that the different
eigenvalues are related to the sets of eigenstates belonging to the different
irreducible representations of the group of symmetry of the hamiltonian.
In particular,  in absence of the time inversion symmetry
violation, the superconducting states relating to 
 the nonequivalent irreducible representations of the point symmetry
group of crystal obey the different critical temperatures.  Similarly  
the eigenstates of the particles in the ferromagnetic crystals  are
classified in accordance with corepresentations $\Gamma$ of magnetic group $M$
of the crystal \cite{13a}.  The latter differ from usual representations
by the law of multiplication of matrices of representation which is
$\Gamma(g_{1}¥)\Gamma(g_{2}¥)=\Gamma(g_{1}¥g_{2}¥)$ for elements $g_{1}¥,
g_{2}¥$ of group $M$ if element $g_{1}¥$ does not include the time
inversion operation and $\Gamma
(g_{1}¥)\Gamma^{*}¥(g_{2}¥)=\Gamma(g_{1}¥g_{2}¥)$ 
if element $g_{1}¥$
does include the time inversion.  The matrices of transformation of
the order parameters (\ref{e9})-(\ref{e12}) by the symmetry operations
of the group $D_{2}(C_{2}^{z})=(E, C_{2}^{z}, RC_{2}^{x},
RC_{2}^{y})$ are just numbers (characters).  As usual for
one-dimensional representations they are equal $\pm 1$.  For the state
$A_{1}¥$ (\ref{e9}) which is a conventional superconducting state
obeying the complete point-magnetric symmetry of initial normal state
they are $(1,1,1,1)$.  For the order parameter $A_{2}¥$ (\ref{e10})
they are $(1,1,-1,-1)$ where $-1$ corresponds to the elements of the
superconducting symmetry class (\ref{e3}) containing the phase factor
$e^{i\pi}¥$.  The same is true for the table of characters of the
other states.  So all the corepresentations in the present case are
real, however their difference from the usual representations manifests
itself in the relationship of equivalence.

The two corepresentations of the group $M$ are called equivalent
\cite{13b} if their matrices $\Gamma(g)$ and $\Gamma'(g)$ are
transformed to each other by means of the unitary matrix $U$ as
$\Gamma'(g)=U^{-1}¥\Gamma(g)U$ if the element $g$ does not include the
time inversion and as $\Gamma'(g)=U^{-1}¥\Gamma(g)U^{*}¥$ if the
element $g$ includes the time inversion.  The corepresentations for
the pair of states $A_{1}¥$ and $A_{2}¥$ are equivalent .  In view of
one-dimensional character of these corepresentations the matrix of the
unitary transformation is simply given by the number $U=i$.  The
states $A_{1}¥$ and $A_{2}¥$ belong to the same corepresentation and
represent two particular forms of the same superconducting state.  It
will be shown below that if we have state $A_{1}¥$ in the ferromagnet
domains with the magnetization directed up the superconducting state
in the domains with down direction of the magnetization corresponds to
the superconducting state $A_{2}¥$.  The same is true for the pair of
states $B_{1}¥$ and $B_{2}¥$.

The critical temperatures of the phase transition from a ferromagnetic
normal state to the superconducting states relating to the
nonequivalent corepresentations are in general different. The
latter is garanteed by the property of the orthogonality of the order
parameters relating to the nonequivalent corepresentations:
\begin{equation}
    \langle {\Psi^{A*}}({\bf k})\Psi^{B}({\bf k}) \rangle=0.
\label{e15}
\end{equation}
 
The critical temperatures of equivalent states $A_{1}¥$ and $A_{2}¥$
in the ferromagnetic domains with the opposite orientations of
magnetization are equal (see below).

 \subsection{Stimulation of superconductivity by ferromagnetism}
 
 All the listed above superconducting phases are in principle
 nonunitary and obey the Cooper pair spin momentum
\begin{equation} 
{\bf S}=i\langle {\bf \Psi}^{*}¥ \times {\bf \Psi}\rangle=
\frac{i}{4}\langle f_{-}¥^{*}¥f_{-}¥-f_{+}¥^{*}¥f_{+}¥\rangle,
\label{e16}
\end{equation}
where $f_{\pm}¥=f_{x}¥\pm if_{y}¥$ and Cooper pair angular momentum
\begin{equation} 
{\bf L}=i\langle {\bf \Psi}^{*}_{i}\left({\bf k}\times
\frac{\partial}{\partial {\bf k}} \right) {\bf \Psi}_{i}\rangle.
\label{e17}
\end{equation}
The spontaneous Cooper pair magnetic moment as it is clear from (\ref{e16})
is proportional to the difference in the density of populations 
of pairs with spin up and spin down.  In superfluid He-3 in the
$A_{1}¥$-state, only Cooper pairs with spin down are present.  Their
magnetic moments interact with external field giving rise to an increase
of the critical temperature of the phase transition to the superfluid
state.  In the ferromagnetic metals with strong spin-orbital coupling
there are the Cooper pairs with any projection of the total spin. 
However the dependence of the critical temperature of the
superconducting phase transition from the ferromagnet magnetization
also exists.  On the microscopic level this dependence originates from
the difference of the pairing interaction and the density of states on
the Fermi surfaces for the particles with opposite spin projections
(see below).  Here one needs to note that as usual the word "spin"
means in fact "pseudospin" and it is used to denote Kramers double
degeneracy of electron states in a metal with spin-orbital coupling.

On the phenomenological level the shift of the critical temperature can be
described by the following term in the Landau free energy expansion
\begin{equation}
    -N_{0}f\left ( \frac{\mu_{b}H}{\varepsilon_{F}¥}\right )|\eta|^{2},
\label{e18}
\end{equation}
where   $N_{0}¥$ is the electron density of states on the Fermi surface,
$\mu_{B}¥$ is the Bohr
magneton,
the function $f(x)\sim x$ at small values of its argument.  The magnetic field
\begin{equation}
{\bf H}={\bf H}_{ex}¥+{\bf H}_{ext}¥
\label{e19}
\end{equation}
consists of the exchange field ${\bf H}_{ex}¥$ and the external magnetic 
field ${\bf H}_{ext}¥$.  Let us stress a very important difference
between these fields.  ${\bf H}_{ex}¥$ is frozen into the crystal. It is
transformed with any operation of the point symmetry group and
completely invariant under magnetic symmetry class $D_{2}¥(C_{2}¥^{z}¥)$
trasformations.  ${\bf H}_{ext}¥$ does not relate to these
transformations, but as any magnetic field, changes the sign under time
inversion.

The exchange field  acting on the electron spins stimulates the
nonunitary superconducting state.  
The resulting  enhancement of the critical temperature can be estimated as
\begin{equation}
   \frac{T_{c}(H_{ex})-T_{c}}{T_{c}} \approx
   \frac{\mu_{B}H_{ex}}{\varepsilon_{F}}.
\label{e20}
\end{equation}
The exchange field determines the relative shift of the Fermi 
surfaces 
for the spin up and spin
down quasiparticules. One can estimate the value of this field
for  $UGe_{2}$ by its Curie temperature.  Taking into account that
at temperatures lower than $\approx 20K$ the phase transition into
ferromagnet state starts to be of first order \cite {14} one can
say that $H_{ex}.$ is lying in the interval $\approx (20T, 40T)$ in
the whole interval of the pressures where superconductivity exists.

Unlike in He-3, in ferromagnetic superconductors the magnetic field acts
through the electron charges on the orbital electron motion to suppress
the superconducting state.  The reduction of the critical temperature
due to the orbital effect is
\begin{equation}
   \frac{T_{c}(H_{em})-T_{c}}{T_{c}}\approx
 -\frac{\xi_{0}^{2}H_{em}}{\Phi_{0}} \approx
 -\frac{\varepsilon_{F}\mu_{B}H_{em}}{T_{c}^{2}}.
\label{e21}
\end{equation}
The electromagnetic field $H_{em}$ acting on the electron charges
is determined by the modulus of the sum of the vectors of the external magnetic
field and the dipole field of its own ferromagnet magnetic moment. 
The latter is much smaller than $H_{ex}$.  In the absence of the
external field one can estimate the value of the $H_{em}$ by the
value of the magnetic moment density which in $UGe_{2}$ is of the order of
1kG \cite {1,2}.

The estimations (\ref{e20}) and (\ref{e21}) shows that the stimulation
of a nonunitary superconductivity by ferromagnetism takes place at
\begin{equation}
   \frac{H_{ex}}{H_{em}}>
   \frac{\varepsilon_{F}^{2}¥}{T_{c}^{2}¥}.
\label{e22}
\end{equation}
The interplay between the effects of the stimulation and the 
suppression of
the critical temperature in ferromagnetic superconductors determines
the phase diagrams of these materials \cite {1,2,3,4}.

One can find the confirmation of these qualitative estimates from
the equation for the critical temperature of the superconducting phase
transion \cite{8} written \footnote{For simplicity we will not use the
comlete form of the equation for the order parameter taking into
account the effect of spontaneous orbital magnetism \cite{15}.} for one of
superconducting state (\ref{e9})- (\ref{e12}) in frame of some
particular model of pairing:

\begin{equation}
\Delta_{\alpha\beta}({\bf R},{\bf r})=-T\sum_{\omega}\int d{\bf r}
V_{\beta \alpha,\lambda\mu}({\bf r},{\bf r'})
G_{\omega}^{\lambda\gamma}({\bf r}) G_{-\omega}^{\mu\delta}({\bf
r})\exp(i{\bf r}{\bf D}({\bf R})) \Delta_{\gamma\delta}({\bf R},{\bf
r}),
\label{e23}
\end{equation}
where
$$
{\bf D}({\bf R})=-i\frac{\partial}{\partial{\bf R}}+\frac{2e}{c}{\bf 
A}({\bf R}),
$$
$$
\Delta_{\alpha\beta}({\bf R},{\bf r})={\bf d}({\bf R},{\hat {\bf r}})
{\bf g}_{\alpha \beta},
$$

\begin{equation}
V^{\Gamma}_{\beta \alpha,\lambda\mu}(\hat {{\bf r}},\hat{\bf r'})
=-\frac{V}{2} ({\bf \Phi}^{\Gamma}(\hat {\bf r}){\bf g}_{\beta 
\alpha}) ({\bf
{\Phi}^{\Gamma}}^{*}¥(\hat {\bf r'}){\bf g}^{\dag}_{\lambda \mu}),
\label{e24}
\end{equation}
${\bf g}_{\alpha \beta}.=i(\mbox {\boldmath $\sigma
$}\sigma_{y})_{\alpha\beta}$, $\mbox {\boldmath
$\sigma$}=(\sigma_{x}, \sigma_{y}, \sigma_{z})$ are the Pauli 
matrices. The functions
\begin{equation}
{\bf \Phi}^{\Gamma}(\hat {\bf r})
=\hat {\bf x}g^{\Gamma}_{x}( \hat{\bf r})+ \hat {\bf
y}g^{\Gamma}_{y}(\hat{\bf r}) + \hat {\bf z}g^{\Gamma}_{z}(\hat 
{\bf r}) 
\label{e25}
\end{equation}
are related to the particular corepresentation and in its particular
form ${\bf \Psi}^{\Gamma}$ (\ref{e9})-(\ref{e12}).  For instanse for
$A_{1}¥$ case it is
\begin{equation}
\Phi^{A_{1}¥}(\hat{\bf r})= \hat {\bf
x}({\hat r}_{x}v_{1}^{A_{1}¥}+i{\hat r}_{y}v_{2}^{A_{1}¥})+ \hat {\bf
y}({\hat r}_{y}v_{3}^{A_{1}¥}+i{\hat r}_{x}v_{4}^{A_{1}¥})+ \hat {\bf
z}({\hat r}_{z}v_{5}^{A_{1}¥}+i{\hat r}_{x}{\hat r}_{y}{\hat 
r}_{z}v_{6}^{A_{1}¥}),
\label{e26}
\end{equation}
where $v_{1}^{A}, \ldots$ are real functions of ${\hat r}_{x}^{2}, {\hat
r}_{y}^{2}, {\hat r}_{z}^{2}$. The normal metal
electron Green functions are diagonal $2\times 2$ matrices
\begin{equation}
G_{\omega}^{\lambda\gamma}({\bf r})=\int \frac{d{\bf 
p}}{(2\pi)^{3}}
e^{i{\bf p}{\bf r}}((i\omega -\xi({\bf p}))\sigma_{0}+
2g({\bf p})\mu_{B}\sigma_{z}H_{ex})^{-1}_{\lambda\gamma}.
\label{e27}
\end{equation}
It is convenient to work with them by introducing the following 
notations
$$
{\hat G}_{\omega}({\bf r})=\frac{1}{2}[(G_{\omega}^{\uparrow}({\bf 
r})
+G_{\omega}^{\downarrow}({\bf 
r}))\sigma_{0}+(G_{\omega}^{\uparrow}({\bf r})
-G_{\omega}^{\downarrow}({\bf r}))\sigma_{z}].
$$

Using the general form of self-consistency equation (\ref{e23}) one can
easily obtain 
the following system of equations
\begin{equation}
\langle g_{-}^{*}({\hat {\bf r}})f_{-}({\hat {\bf 
r}})\rangle\eta({\bf R})=
\langle g_{-}^{*}({\hat {\bf r}})g_{-}({\hat {\bf r}})\rangle{\hat
L}\eta({\bf R}),
\label{e28a}
\end{equation}
\begin{equation}
\langle g_{z}^{*}({\hat {\bf r}})f_{z}({\hat {\bf 
r}})\rangle\eta({\bf R})=
\langle g_{z}^{*}({\hat {\bf r}})g_{z}({\hat {\bf r}})\rangle{\hat
L}\eta({\bf R}),
\label{e28b}
\end{equation}    
\begin{equation}
\langle g_{+}^{*}({\hat {\bf r}})f_{+}({\hat {\bf 
r}})\rangle\eta({\bf R})=
\langle g_{+}^{*}({\hat {\bf r}})g_{+}({\hat {\bf r}})\rangle{\hat
L}\eta({\bf R}),
\label{e28c}
\end{equation}
where the combinations $ g_{\pm}=g_{x}\pm ig_{y}, ~~g_{z}$ and
$f_{\pm}=f_{x}\pm if_{y}, ~~f_{z}$
correspond to the pairing interaction and the order parameter amplitudes
with spins up-up, down-down and zero projection of the pair spin on the
$z$-direction.  The angular brackets denote the averaging over the
directions of unit vector $\hat {\bf r}$ and the integral operator in
the right hand side is
\begin{eqnarray}
{\hat L}\eta({\bf R})&=&\frac{1}{2}VT\sum_{\omega}\int d{\bf
r}[g_{-}^{*}({\hat {\bf r}})f_{-}({\hat {\bf r}})
G_{\omega}^{\uparrow}({\bf r})G_{-\omega}^{\uparrow}({\bf
r}) +g_{z}^{*}({\hat {\bf r}})f_{z}({\hat {\bf 
r}})(G_{\omega}^{\uparrow}({\bf
r})G_{-\omega}^{\downarrow}({\bf r})+G_{\omega}^{\downarrow}({\bf
r})G_{-\omega}^{\uparrow}({\bf r})) \nonumber \\
&+&g_{+}^{*}({\hat {\bf r}})f_{+}({\hat {\bf r}})
G_{\omega}^{\downarrow}({\bf r})G_{-\omega}^{\downarrow}({\bf
r}) ] \exp(i{\bf r}{\bf D}({\bf R}))\eta({\bf R}).
\label{e29}
\end{eqnarray}
One must to add to this equation the
normalization condition (\ref{e8})
\begin{equation}
\langle\frac{1}{2}(f_{+}^{*}({\hat {\bf r}})f_{+}({\hat {\bf r}})+
f_{-}^{*}({\hat {\bf r}})f_{-}({\hat {\bf r}}))+
f_{z}^{*}.({\hat {\bf r}})f_{z}({\hat {\bf r}})\rangle=1.
\label{e30}
\end{equation}
After finding of the eigen function $\eta({\ R})$ of the operator 
$\hat L$
one must find the critical temperature from the condition of zero
value of the determinant of the linear system of equations
(\ref{e28a})-(\ref{e28c}) for the amplitudes $\langle g^{*}f \rangle$.
It is worth noting that, for the interaction in the form of $A_{1}¥$
state, the order parameter can be chosen correspondingly as belonging to
one of $A_{1}¥$ or $A_{2}¥$ states.  As a result all the amplitudes
$\langle g^{*}f \rangle$ will be correspondingly real or imaginary and
we deal with the system of equations of the third order.  The same is
correct for the $B$ type corepresentation.

Then the appearence of the linear shift of the critical temperature
due to exchange field (\ref{e20}) follows trivially from the linear
shifts of the amplitudes
$$
\langle g_{-}^{*}g_{-}\rangle-\langle g_{-}^{*}g_{-}\rangle({H}¥_{ex}¥=0)
\sim -\frac{\mu_{B}¥H_{ex}¥}{\varepsilon_{F}¥},
$$
$$
\langle g_{+}^{*}g_{+}\rangle-\langle g_{+}^{*}g_{+}\rangle({H}¥_{ex}¥=0)
\sim \frac{\mu_{B}¥H_{ex}¥}{\varepsilon_{F}¥},
$$
$$
\int d{\bf r}(G_{\omega}^{\uparrow}G_{-\omega}^{\uparrow}-
G_{\omega}^{\uparrow}(H_{ex}¥=0)G_{-\omega}^{\uparrow}(H_{ex}¥=0))
\sim-\frac{\mu_{B}¥H_{ex}¥}{\varepsilon_{F}¥},
$$
$$
\int d{\bf r}(G_{\omega}^{\downarrow}G_{-\omega}^{\downarrow}-
G_{\omega}^{\downarrow}(H_{ex}¥=0)G_{-\omega}^{\downarrow}(H_{ex}¥=0))
\sim\frac{\mu_{B}¥H_{ex}¥}{\varepsilon_{F}¥}.
$$
To demonstrate the validity of two latter relationships it is enough
to look at the expression for electron Green function in a normal metal with
isotropic spectrum $\xi({\bf p})=\xi $ and isotropic g-factor $g({\bf
p})=1/2$ (see \cite {15}):
\begin{equation}
G_{\omega}^{\lambda
}({\bf r})=-\frac{\pi N_{0}}{p_{0} r}
\exp\left (ip_{0}^{\lambda} r sign\omega 
-\frac{r|\omega|}{v_{0}^{\lambda}}
\right ),
\label{e31}
\end{equation}
here 
$p_{0}^{\lambda}=[2m(\varepsilon_{F}-\lambda\mu_{B}H_{ex})]^{1/2}$,

$p_{0}=p_{0}^{\lambda}(H_{ex}=0)$ is the Fermi momentum,
$\lambda=\uparrow, \downarrow$ or $+1,-1$, $v_{0}^{\lambda}=
p_{0}^{\lambda}/m$, is the Fermi velocity on the corresponding sheet
of the Fermi surface, $v_{0}=p_{0}/m$, $N_{0}mp_{0}/2\pi^{2}$ is
the density of states on the one spin projection. 

As for the second term in the right hand side of the equation (\ref{e29})
due to the difference in the Fermi momenta with spin up and spin down
it contains the fast oscillating products of two Green functions and
starts to be negligibly small.  The smallness of this term however does
not result in the disappearence of the amplitude $f_{z}$ of the Cooper
pair state with zero projection of spin becouse all three amplitudes
$f_{+},f_{-}, f_{z}$ obey coupled linear equations
(\ref{e28a})-(\ref{e28c}).  This fact is the direct consequence of the
strong spin-orbital coupling.  Unlike this, in the superfluid He-3,
all three amplitudes $f_{+},f_{-}, f_{z}$ obey independent equations
characterized by different critical temperatures \cite{15} such that
the amplitudes $f_{+}¥$ and $f_{z}$ are equal to zero at the critical
temperature where the amplitude $f_{-}¥$ appears.

Generally speaking the second term in the (\ref{e29}) promotes the
appearance of the oscillating solution
\begin{equation}
\eta({\bf R})=\eta(x,y)e^{iQz}.
\label{e32}
\end{equation}
On the other hand the first and the third terms in the equation 
(\ref{e29})
make these oscillations nonprofitable (oscillations in the order
parameter decrease the critical temperature).  In  superconductors
with s-pairing the appearence of a solution with nonvanishing $Q$ or so
called Fulde-Ferrell-Larkin-Ovchinnikov state \cite{16,17} is possible
for  large enough values of $H_{c2}^{orb}$ in comparison with
the paramagnetic limiting field \cite{18}.  In superconductors with
triplet pairing when $H_{ex}\gg H_{c2}^{orb}$ one would not expect the
appearance of FFLO state.  This question however demands a special
investigation in the frame of some particular model of pairing
interaction.

\subsection{Domain structures}

Let us assume now that we have interactions (\ref{e24})-(\ref{e25}) in the form
corresponding to $A_{1}¥$ and $A_{2}¥$ states  such that the
corresponding functions $v_{i}¥^{A}¥_{1}¥$ and $v_{i}¥^{A}¥_{2}¥$ are equal.
Let us fixe the solutions of the two corresponding sets of equations
(\ref{e28a})-(\ref{e28c}) as relating to $A_{1}¥$ and $A_{2}¥$ states.
\footnote{As it has been mentioned above one can discuss also 
another pair of solutions relating correspondingly to $A_{2}¥$ and
$A_{1}¥$ states.  This choice does not change the conclusions of this
subsection.} Such a pair of states possess equal and opposite
direction Cooper pair magnetic moments.  It is easy to see that the
following equalities are obeyed
$$\langle g_{-}¥^{A_{1}*¥}¥g^{A_{1}¥}_{-}¥
\rangle=\langle g_{+}¥^{A_{2}*¥}¥g^{A_{2}¥}_{+}¥ \rangle,~~ \langle
g_{+}¥^{A_{1}*¥}¥g^{A_{1}¥}_{+}¥ \rangle=\langle
g_{-}¥^{A_{2}*¥}¥g^{A_{2}¥}_{-}¥ \rangle,~~ \langle
g_{z}¥^{A_{1}*¥}¥g^{A_{1}¥}_{z}¥ \rangle=\langle
g_{z}¥^{A_{2}*¥}¥g^{A_{2}¥}_{z}¥ \rangle.$$
Hence, if the state $A_{1}¥$ is the solution of the system of
equations (\ref{e28a})-(\ref{e28c}) with critical temperature
$T_{c}¥$, the state $A_{2}¥$ is also the solution of the system
(\ref{e28a})-(\ref{e28c} with opposite direction of the $H_{ex}¥$ and
the same critical temperature.  This means, if the pairing interaction
in the ferromagnet with up direction of $H_{ex}¥$ corresponds to the
pure $A_{1}¥$ state, the superconducting states in ferromagnetic
domains with opposite orientation of magnetization will be $A_{2}¥$. 
One can say that this is the consequence of the above mentioned property of
conjugacy between the states $A_{1}¥$ and $A_{2}¥$.  The same is true
for another pair of conjugate states $B_{1}¥$ and $B_{2}¥$.

The ferromagnet domain structure with
alternating up-down direction of the magnetization is always
accompanied by the superconducting domain structure with alternating
properties of the complex conjugacy of the order parameter and
alternating up-down direction of the Cooper pair magnetic moment.  The
superconducting order parameter distribution in the vicinity of the
domain wall between of two adjacent domains demands special
investigation.
 
 It is quite natural that the Abrikosov vortices having in the $A_{1}¥$-state
 some fixed direction of the current and flux will have  opposite
 orientations of the current and flux in the adjacent ferromagnet
 domain with the opposite direction of magnetization.

\section{Superconductivity in ferromagnet metals with cubic symmetry}

Symmetric orientations of magnetic moments in cubic crystals along the
symmetry axes of the fourth or of the third order give rise to a
decreasing of the initial cubic symmetry of the normal state to the
magnetic classes $D_4(C_4) =(E, C_4, C_2, C_4^3, RU_x, RU_y, RU',
RU'',)$ and $D_3(C_3) =(E, C_3, C_3^2, RU_1, RU_2, RU_3)$
correspondingly \cite{13}.  Let as look first at the tetragonal
magnetic class.

\subsection{Tetragonal magnetic class $D_4(C_4)$}

As before we construct first the groups being isomorphic to the
initial magnetic group $D_4(C_4)$
by means of combining  its elements with $e^{i\pi}¥$ and $e^{\pm
i\pi/2}¥$ phase factors from the group of the gauge transformations
$U(1)$.  The explicit form of these superconducting magnetic classes
are
\begin{equation}
D_4(C_4) =(E, C_4, C_2, C_4^3, RU_x, RU_y, RU', RU''),
\label{e33}
\end{equation}
\begin{equation}
\tilde D_4(C_4) =(E, C_4, C_2, C_4^3, e^{i\pi}RU_x, e^{i\pi}RU_y, 
e^{i\pi}RU', e^{i\pi}RU''),
\label{e34}
\end{equation}
\begin{equation}
\tilde D_4(D_2) =(E, e^{i\pi}C_4, C_2, e^{i\pi}C_4^3, RU_x, RU_y, 
e^{i\pi}RU', e^{i\pi}RU''),
\label{e35}
\end{equation}
\begin{equation}
\tilde D_4(D'_2) =(E, e^{i\pi}C_4, C_2, e^{i\pi}C_4^3, e^{i\pi}RU_x, 
e^{i\pi}RU_y, RU', RU''),
\label{e36}
\end{equation}
\begin{equation}
D_4(E) =(E, e^{i\pi/2}C_4, e^{i\pi}C_2, e^{3i\pi/2}C_4^3, e^{i\pi}RU_x, 
RU_y, e^{3i\pi/2}RU', e^{i\pi/2}RU''),
\label{e37}
\end{equation}
\begin{equation}
D_4(E) =(E, e^{3i\pi/2}C_4, e^{i\pi}C_2, e^{i\pi/2}C_4^3, e^{i\pi}RU_x, 
RU_y, e^{i\pi/2}RU', e^{3i\pi/2}RU'').
\label{e38}
\end{equation}

The order parameters of the superconducting states corresponding to 
these classes are
\begin{equation}
\Psi^{A_1}({\bf k})= \hat {\bf 
x}(k_{x}u_{1}^{A_1}-ik_{y}u_{2}^{A_1})+ \hat
{\bf y}(k_{y}u_{1}^{A_1}+ik_{x}u_{2}^{A_1})+
\hat {\bf z}(k_{z}u_{3}^{A_1}+ik_{x}k_{y}k_{z} 
(k_{x}^2-k_{y}^2)u_{4}^{A_1}),
\label{e39}
\end{equation}
\begin{equation}
\Psi^{A_2}({\bf k})= \hat {\bf 
x}(ik_{x}u_{1}^{A_2}-k_{y}u_{2}^{A_2})+ \hat
{\bf y}(ik_{y}u_{1}^{A_2}+k_{x}u_{2}^{A_2})+
\hat {\bf z}(ik_{z}u_{3}^{A_2}+k_{x}k_{y}k_{z} 
(k_{x}^2-k_{y}^2)u_{4}^{A_2}),
\label{e40}
\end{equation}
\begin{equation}
\Psi^{B_1}({\bf k})= \hat {\bf 
x}(k_{x}u_{1}^{B_1}+ik_{y}u_{2}^{B_1})+ \hat
{\bf y}(-k_{y}u_{1}^{B_1}+ik_{x}u_{2}^{B_1})+
\hat {\bf 
z}(k_{z}(k_{x}^2-k_{y}^2)u_{3}^{B_1}+ik_{x}k_{y}k_{z} 
u_{4}^{B_1}),
\label{e41}
\end{equation}
\begin{equation}
\Psi^{B_2}({\bf k})= \hat {\bf 
x}(ik_{x}u_{1}^{B_2}+k_{y}u_{2}^{B_2})+ \hat
{\bf y}(-ik_{y}u_{1}^{B_2}+k_{x}u_{2}^{B_2})+
\hat {\bf 
z}(ik_{z}(k_{x}^2-k_{y}^2)u_{3}^{B_2}+k_{x}k_{y}k_{z} 
u_{4}^{B_2}),
\label{e42}
\end{equation}
\begin{equation}
{\bf \Psi}^{E_{+}¥}¥({\bf k})=(k_{x}¥+ik_{y}¥)[{\hat z}u_{1}^{E_{+}¥}¥+
ik_z(k_x{\hat y} -k_y{\hat x})u_{2}¥^{E_{+}¥}¥]+ ({\hat x}+i{\hat y})
[k_{z}¥u_{3}¥^{E_{+}¥}¥+ ik_{x}¥k_{y}¥k_{z}¥(k_{x}^2-k_{y}^2)
u_{4}¥^{E_{+}¥}¥],
\label{e43}
\end{equation}
\begin{equation}
{\bf \Psi}^{E_{-}¥}¥({\bf k})=(k_{x}¥-ik_{y}¥)[{\hat z}u_{1}^{E_{-}¥}¥+
ik_z(k_x{\hat y} -k_y{\hat x})u_{2}¥^{E_{-}¥}¥]+ ({\hat x}-i{\hat y})
[k_{z}¥u_{3}¥^{E_{-}¥}¥+ ik_{x}¥k_{y}¥k_{z}¥(k_{x}^2-k_{y}^2)u_{4}¥^{E_{-}¥}¥],
\label{e44}
\end{equation}
where $u_{1}^{A_1}, \ldots$ are real functions of
$k_{x}^{2}+k_{y}^{2}, k_{z}^{2}$.

As for the orthorombic case the states $A_{1}¥$, $A_{2}¥$ and $B_{1}¥$,
$B_{2}¥$ represent the pairs of equivalent corepresentations.  Another
two superconducting states $E_{+}¥$ and $E_{-}¥$ are related to
nonequivalent corepresentations.  In total there are four different
superconducting states.  The states $A$ and $E_{\pm}¥$ have no
symmetry zeros in the quasiparticle spectra.  Only the states of $B$
type have symmetry points of zeros lying on the nothern and southern
poles of the Fermi surface $k_x=k_y=0$.  This is easy to see directly
from the expressions (\ref{e39})-(\ref{e44}).

Again in alternating ferromagnet domains with  opposite directions of
the magnetization there is  alternating sequence of $A_{1}¥$ and
$A_{2}¥$, or $B_{1}¥$ and $B_{2}¥$, or $E_{+}¥$ and $E_{-}¥$states. 
As for the latter pair of states one can check this statement directly
from the system of equations (\ref{e28a})-(\ref{e28c}).

\subsection{Trigonal magnetic class $D_3(C_3)$}

 The groups being isomorphic to to
the initial magnetic group $D_{3}¥(C_{3}¥)$ are constructed by the
combinations of its elements with elements $e^{i\pi}¥$ and $e^{\pm
2i\pi/3}¥$ of the gauge group $U(1)$.  That yields the 
superconducting magnetic classes of symmetry 
\begin{equation}
D_3(C_3)=(E, C_{3}¥, C_{3}¥^{2}¥, RU, RU_{2}¥, RU_{3}¥) ,
\label{e47}
\end{equation}
\begin{equation}
\tilde D_3(C_3)=(E, C_{3}¥, C_{3}¥^{2}¥, e^{i\pi}¥RU_{1}¥,
e^{i\pi}¥RU_{2}¥,e^{i\pi}¥ RU_{3}¥) ,
\label{e48}
\end{equation}
\begin{equation}
D_3(E)=(E, e^{ 2i\pi/3}¥C_{3}¥, e^{- 2i\pi/3}¥C_{3}¥^{2}¥, RU,
e^{- 2i\pi/3}¥RU_{2}¥, e^{ 2i\pi/3}¥RU_{3}¥) ,
\label{e49}
\end{equation}
\begin{equation}
\tilde D_3(E)=(E, e^{- 2i\pi/3}¥C_{3}¥, e^{2i\pi/3}¥C_{3}¥^{2}¥, RU,
e^{ 2i\pi/3}¥RU_{2}¥, e^{-2i\pi/3}¥RU_{3}¥) ,
\label{e50}
\end{equation}
where the elements $U_{1}¥, U_{2}¥, U_{3}¥$ are the rotations on the
angle $\pi$ around axes
$$
{\hat \phi}_{1}¥=\hat {\bf x}, 
~{\hat \phi}_{2}¥=\frac{1}{2}(-\hat {\bf x}+{\sqrt 3}\hat {\bf y}),
~{\hat \phi}_{3}¥=\frac{1}{2}(-\hat {\bf x}-{\sqrt 3}\hat {\bf y}).
$$
 
The corresponding order parameters are
\begin{equation}
\Psi^{A_1}({\bf k})= i(k_{x}¥\hat {\bf y}-k_{y}\hat {\bf
x})u_{1}^{A_1}+ k_{z}¥\hat {\bf z}u_{2}¥^{A_{1}¥}¥+ (k_{y}¥\hat {\bf
y}-k_{x}\hat {\bf x})u_{3}¥^{A_{1}¥}¥
\label{e51}
\end{equation}
\begin{equation}
\Psi^{A_2}({\bf k})= 
(k_{x}¥\hat {\bf y}-k_{y}\hat {\bf x})u_{1}^{A_2}+
ik_{z}¥\hat {\bf z}u_{2}¥^{A_{2}¥}¥+ i(k_{y}¥\hat {\bf y}-k_{x}\hat
{\bf x})u_{3}¥^{A_{2}¥}¥,
\label{e52}
\end{equation}
\begin{eqnarray}
\Psi^{E_{+}¥}¥({\bf k})&=& (\phi_{1}¥ +
e^{ 2i\pi/3}¥\phi_{2}¥+e^{ -2i\pi/3}¥\phi_{3}¥)
[i\hat {\bf z}u_{1}^{E_{+}¥}¥+ k_{z}¥(k_{x}¥\hat {\bf y}-k_{y}¥\hat {\bf
y})u_{2}¥^{E_{+}¥}¥]+ \\ \nonumber
&(&\hat \phi_{1}¥ + e^{ 2i\pi/3}¥\hat
\phi_{2}¥+e^{ -2i\pi/3}¥\hat \phi_{3}¥) [ik_{z}¥u_{3}^{E_{+}¥}¥+
\phi_{1}¥\phi_{2}¥\phi_{3}¥u_{4}¥^{E_{+}¥}¥ ],
\label{e53}
\end{eqnarray}
\begin{eqnarray}
\Psi^{E_{-}¥}¥({\bf k})&=&(\phi_{1}¥ +
e^{- 2i\pi/3}¥\phi_{2}¥+e^{ 2i\pi/3}¥\phi_{3}¥)
[i\hat {\bf z}u_{1}^{E_{-}¥}¥+ k_{z}¥(k_{x}¥\hat {\bf y}-k_{y}¥\hat {\bf
y})u_{2}¥^{E_{-}¥}¥] + \\ \nonumber
&(&\hat \phi_{1}¥ + e^{- 2i\pi/3}¥\hat
\phi_{2}¥+e^{ 2i\pi/3}¥\hat \phi_{3}¥) [ik_{z}¥u_{3}^{E_{-}¥}¥+
\phi_{1}¥\phi_{2}¥\phi_{3}¥u_{4}¥^{E_{-}¥}¥ ],
\label{e54}
\end{eqnarray}
where
$$
{\phi}_{1}¥=k_{x}¥, ~{\phi}_{2}¥=\frac{1}{2}(-k_{x}¥+
{\sqrt 3}k_{y}¥), ~{\phi}_{3}¥=\frac{1}{2}(-k_{x}¥-{\sqrt 3}k_{y}¥)
$$
and $u_{1}^{A_1}, \ldots$ are the real functions invariant under the
transformations $D_{3}¥$ group.  

As before the $A_{1}¥$ and $A_{2}¥$ states correspond to the equivalent
corepresentations. The states $E_{\pm}¥$ are related to nonequivalent representations.

None of these states have the symmetry nodes in the quasiparticle spectra.

\section{Conclusion}

The symmetry classifications of the superconducting states 
with triplet pairing in the orthorombic and cubic ferromagnet crystals with
strong spin-orbital coupling is presented.  It is found that unlike 
the case of weak spin-orbital interaction where the nonunitary
magnetic superconducting states are possible only in the case of
multicomponent superconductivity \cite{8} any superconducting state in the
ferromagnet metals with strong spin-orbital coupling is in general nonunitary.

The ferromagnetism stimulates in general the triplet superconductivity
even with a one-component order parameter.  The mechanism of this stimulation is
due to the difference of the pairing interaction and the density of states for
electrons with opposite directions of spin of degree of ferromagnet
Fermi-liquid polarization.  However the competitive mechanism
supressing  superconductivity due to the orbital diamagnetic currents
is always presents.  The interplay between these two interactions
determines the superconducting phase diagram in the metallic
ferromagnet materials.

The presence of the ferromagnet domain structure in the
superconducting state is always accompanied by the corresponding
superconducting domain structure of the complex conjugate states.  The
adjacent domains contain the quantized vortices with opposite
directions of currents and fluxes.  

\section{Acknowledgments}
I am grateful to I.A.Fomin and A.D.Huxley for the numerous wholesome
discussions and D. Braithwaite for the valuable help.  In particular I
would like to acknowledge the strong influence of K.V.Samokhin in
reaching the present formulation.



\end{document}